\documentstyle{article}
\makeatletter
\def\@equaligncr{\cr \noalign{\vskip 5pt}
                 \vphantom{\displaystyle\int\limits^1_{-1}}}
\long\def\equalign #1\end#2{
    \let \\=\@equaligncr
    \refstepcounter{equation}$$ \vcenter{\tabskip=5pt
    \halign{\hfil$\displaystyle{##}$&&$\displaystyle{{}##}$\hfil\cr
    \vphantom{\displaystyle\int\limits^1_{-1}}
    #1 \cr}}\end{#2}}

\def\fmref #1{(\ref{#1})}
\def\lds{|\mkern-2.5mu|}
\def\rds{|\mkern-2.5mu|}
\def\ldb{\langle\mkern-4mu\langle}
\def\rdb{\rangle\mkern-4mu\rangle}

\def\Vacket{\lds 0,0\rdb}
\def\Vacbra{\ldb 0,0\rds}
\def\RGV{\lds O(\beta) \rdb}
\def\LGV{\ldb O(\beta) \rds}
\def\Av#1{\left\langle #1\vphantom{\int}\right\rangle}
\def\td{\tau_3}

\def\B{{\cal B}}
\def\Bi{\B^{-1}}

\def\ankh{{\dagger\mkern-2mu\dagger}}
\def\wtilde#1{\widetilde{#1}}
\def\a{a}
\def\at{\wtilde{a}}
\def\ad{a^\dagger}
\def\atd{\wtilde{a}^\dagger}
\def\x{\xi}
\def\xt{\wtilde{\xi}}
\def\xd{\xi^{\ankh}}
\def\xtd{\wtilde{\xi}^{\ankh}}
\def\bk{{\bf k}}

\def\bx{{\bf x}}

\def\ee#1{\mbox{e}^{#1}}

\def\i{{\mbox i}}
\def\ie{\mbox{\small i}}

\begin{document}
\title{Diagonalization of full finite temperature Green's functions\\
       by quasi-particles}
\author{P.A.Henning\thanks{
        E-mail address: phenning@rzhp9a.gsi.de}\\
        Institut f\"ur Kernphysik der TH Darmstadt and GSI\\
        P.O.Box 110552, D-6100 Darmstadt, Germany\\
        and H.Umezawa\thanks{
        E-mail address: umwa@phys.ualberta.ca}\\
        The Theoretical Physics Institute, Univ. of Alberta\\
        Edmonton, Alberta T6G 2J1, Canada}
\date{Phys.Lett. B in Press, 1993}
\maketitle
\begin{abstract}
For thermal systems, standard perturbation theory breaks
down because of the absence of stable, observable asymptotic states.
We show, how the introduction of {\it statistical}
quasi-particles (stable, but not observable) gives rise to a
consistent description. Statistical and spectral information
can be cleanly separated also for interacting systems.
\end{abstract}
\centerline{PACS No. 03.70.+k,05.30.-d,05.20.Dd,05.60.+w,05.70.Ln}
\clearpage
%%%%%%%%%%%%%%%%%%%%%%%%%%%%%%%%%%%%%%%%%%%%%%%%%%%%%%%%%%%%%%%%%%%%%%%%%%%%%%%
\section{Introduction}
It is a well established fact by now, that naive perturbation
theory breaks down at finite temperature, see refs.
\cite{K89,E90} for a recent overview.  This is due the
modification of space-time symmetry in the presence of matter or
a heat bath \cite{BS75}, i.e.  to the absence of stable
asymptotic states for the observable physical particles. To
overcome this problem, first of all one has to employ a
''doubling'' of the Hilbert space, leading to two-point Green
functions that are $2\times 2$ matrices.

The Schwinger-Keldysh \cite{SKM} or Closed-Time-Path method (CTP)
is such a model, which has been used to derive transport
equations, e.g. for nuclear phenomena, over several decades
\cite{KB62,DP91}. However, the physical interpretation of the
matrix structure of the propagator remains quite obscure in this
formalism, and no justification is obtained for a perturbation
expansion.

Another model is called Thermo Field Dynamics (TFD)
\cite{AU87,Ubook}.  It differs from CTP in the important aspect,
that the doubling of the Hilbert space is due to two disjoint,
mutually (anti-) commuting representations of the canonical
commutation relations. Thus, in contrast to CTP, TFD has a
firm mathematical basis \cite{HHW67} and can be used perturbatively.
To this end, one has to sacrifice either the stability aspect or the
observability aspect of asymptotic states. The first
leads to a perturbation expansion in terms of unstable particles
\cite{L88}. It is the purpose of the present paper to
demonstrate, that the second way leads to a simpler and
straightforward approach.

To this end we show, that the thermal instability
of observable states
can be absorbed into a Bogoliubov transformation also
for interacting systems. This Bogoliubov transformation
then defines stable, albeit non-observable, quasi-particles
which serve as basis for a perturbation expansion \cite{HU92}.
Throughout this paper we term them, for historical more than
physical reasons, {\it statistical} quasi-particles \cite{BD71}.
%%%%%%%%%%%%%%%%%%%%%%%%%%%%%%%%%%%%%%%%%%%%%%%%%%%%%%%%%%%%%%%%%%%%
\section{Thermal Bogoliubov transformation}
To establish the notation, we first discuss TFD for a single bosonic
quantum state. We introduce creation and annihilation operators
$\ad$, $\a$, $\atd$, $\at$ for the two representations,
with canonical commutation relations
\begin{equation}
\left[\a,\ad\right] = 1 \;\;\;\;\;\;
\left[\at,\atd\right] =1\;\;\;\;\;\;
\left[\a,\atd\right] =0
\;\end{equation}
(see ref. \cite{Ubook} for a complete discussion).
The $\a$, $\at$ operators annihilate
the physical vacuum $\Vacket$and the two sets
are transformed into another by means of an anti-unitary
mapping, called the {\it tilde} conjugation (see \cite{EHUY91}
for the tilde conjugation rules). For brevity, we work in the
$\alpha=1$ representation, i.e., the thermal equilibrium state
at inverse temperature $\beta$ and chemical potential $\mu$
is described by two state vectors
\begin{equalign}
\RGV& =&\exp\left( f \ad\atd \right)\, \Vacket
\;\;\;\;\; f = \ee{-\beta(\omega-\mu)}\\
\LGV& =&\Vacbra\,\exp\left( \a\at \right)
\; \end{equalign}
Within this framework, the ensemble average of an observable
${\cal E}[\a,\ad]$
is calculated as the expectation value
\begin{equation}\label{av3}
\Av{ {\cal E} } =
\frac{ \LGV \;{\cal E} \;\RGV}{\LGV  O(\beta)\rdb}
\;.\end{equation}
Obviously then the state vectors $\LGV$ and $\RGV$ are annihilated by
certain linear combinations of the
above operators,
\begin{equation}\label{tsc}
\x\RGV=0\;\;\;\;\xt\RGV=0 \;\;\;\;\LGV\xd=0 \;\;\;\;\LGV\xtd=0
\;, \end{equation}
obtained as
\begin{equation}\label{bdef}
\left(\array{r}\x\\
\xtd\endarray\right)=
  \B
  \left(\array{r}\a\\ \atd\endarray\right)
 \;\;\;\;
\left(\array{r}\xd\\ -\xt\endarray\right)^T=
  \left(\array{r}\ad\\-\at\endarray\right)^T
  \Bi
\;.\end{equation}
$\B$ is a 2$\times$2 matrix with determinant 1. Since the $\x$-operators
obey similar commutation relations as do the $\a$-operators,
they define our quasi-particles.
It is crucial to realize, that these entities have nothing
in common with the ''usual'' definition of quasi-particles,
which refers to physical states with an almost pointlike mass
spectrum.

Eqn. \fmref{bdef} is essentially a Bogoliubov transformation
\cite{h90ber}.  The most general form for the Bogoliubov matrix
compatible with our choice of state vectors is then
\begin{equation}\label{bform}
\B = \frac{1}{\sqrt{1- f}}\exp\left(s\tau_3\right)
\left(\array{cc}1 & -\;f\\
                   -1 & 1\endarray \right)
\;,\end{equation}
where $\td=\mbox{diag}(1,-1)$ is a Pauli matrix, $s$ is
free parameter and $f$ is the statistical weight of a
physical single particle state within the ensemble. Therefore the
number density of the physical particles is
\begin{equation}\label{nd}
n = \frac{f}{1-f}
\;.\end{equation}
We find, that the following considerations
are extremely simplified by chosing
$s = 1/2\,\log(1+n) =-1/2\, \log(1-f)$,
since then the Bogoliubov transformation is linear in the density
parameter
\begin{equation}\label{lc}
\B =
\left(\array{cc}1+n & -n\\
                -1 & 1\endarray\right)
\;.\end{equation}
%%%%%%%%%%%%%%%%%%%%%%%%%%%%%%%%%%%%%%%%%%%%%%%%%%%%%%%%%%%%%%%%%%%%
\section{Interacting Boson field}
We now consider a fully interacting bosonic
quantum field. At finite temperature the irreducible
representations of the space-time symmetry group are characterized
by two rather than one continuous parameter \cite{BS75,L88}.
Hence, the {\it interacting} field can be expand into
modes with definite energy and momentum as
\begin{equation}
\phi(x) =  \int\limits_0^\infty\!\! dE \,
           \int\!\!\frac{d^3\bk}{\sqrt{(2\pi)^3}}\,
           \rho^{1/2}(E,\bk)\,\phi_{E,\bk}(x)
\;\end{equation}
The commutation relation of these fields are, in general, not known.
However, we want to calculate only the  two-point Green function
of the interacting field, i.e. it is sufficient to know the
expectation value of the commutator of these fields.

This expectation value in turn can be absorbed in the definition
of the weight function $\rho(E,\bk)$, i.e. for the field operators
we can define
\begin{equation}
\Av{ \left[ \phi_{E,\bk}(t,\bx),\partial_t
            \phi_{E^\prime,\bk^\prime}(t,\bx^\prime) \right] }
 = \ee{\i \bk(\bx-\bx^\prime)}\,2E\,\delta(E-E^\prime)
 \,\delta(\bk-\bk^\prime)
\;.\end{equation}
In other words, for the calculation of bilinear expectation values
of interacting fields it is
sufficient to consider the $\phi_{E,\bk}(x)$ as generalized
free fields \cite{L88}. The full information about the
single-particle spectrum of the theory is contained in the
weight function $\rho(E,\bk)$, and we require the
normalization
\begin{equation}\label{norm}
\int\!\!dE^2 \, \rho(E,\bk) =1
\;.\end{equation}
Note, that the existence of this spectral
decomposition is only guaranteed in case the system is space-time
translation invariant, i.e., if it is in a thermal equilibrium
state.

We now apply the thermal quasi-particle concept to each
energy-momentum eigenmode of the system, i.e. we define
quasi-particle operators $\x_{E,\bk}$
associated with energy-momentum
eigenstates and annihilating the statistical state vectors as
in \fmref{tsc}. These are then Bogoliubov transformed
into physical particle operators for definite energy and momentum,
and those are summed with the above weight function to
operators $\a_k(t)$ such that the interacting field
is
\begin{equation}
\phi(x) =  \int\!\!\frac{d^3\bk}{\sqrt{(2\pi)^3}}\,
                              \left(\ad_k(t)\ee{-\ie\bk\bx}
                                   +\a_k(t)\ee{\ie\bk\bx}\right)
\; \end{equation}
(note the absorption of the usual energy normalization factor
into the operators). In this expansion
of course, the ''creation'' and ''annihilation'' operators
are quite complex objects. From the above reasoning, i.e.
their decomposition into the modes of definite energy and momentum,
we then obtain
\begin{equalign}\label{sdef}
\LGV \a_k^{\dagger(a)}(t) &\a_{k^\prime}^{(b)}(t^\prime)\RGV\\
   =&\delta^3(\bk-\bk^\prime)\,
  \int\!\!dE\,\rho(E,\bk)\;
  \left(\td\B^T(n_{E\bk})\td\right)_{a2}
\left(\B^T(n_{E\bk})\right)^{-1}_{2b}
  \ee{\ie E(t-t^\prime)}\\
\LGV \a_k^{(a)}(t) &\a_{k^\prime}^{\dagger(b)}(t^\prime)\RGV\\
   =&\delta^3(\bk-\bk^\prime)\,
     \int\!\!dE\,\rho(E,\bk)\;
   \left(\B(n_{E\bk})\right)^{-1}_{a1}
\left(\td\B(n_{E\bk})\td\right)_{1b}
   \ee{-\ie E(t-t^\prime)}
\;.\end{equalign}
The parameter $n_{E\bk}$ defining the individual Bogoliubov
transformations thus appears under the energy integral.
While this result can be understood intuitively, i.e.
every mode is in thermal equilibrium with the rest of the
system, it can also be understood in a more formal way.

To this end on has to look at the time evolution of
initially free particle operators: it is highly nonlinear.
Therefore the Bogoliubov transformation in interacting systems
is non-linear, and this non-linearity is reflected in the above
energy integral. It is indeed possible to derive the above
equation without ever touching the concept of generalized
free fields \cite{HU92}.

Note, that in our notation the weight function $\rho$ has only support
for positive energy arguments. The retarded and advanced propagator are
in momentum space
\begin{equation}\label{rap}
D^{R,A}(E,\bk)  =
  \int\!\!dE^\prime\;\rho(E^\prime,\bk)\;
   \left(\frac{1}{E-E^\prime\pm\i\epsilon}
        -\frac{1}{E+E^\prime\pm\i\epsilon} \right)
\;,\end{equation}
and the limit of free particles is recovered when
\begin{equation}\label{fb}
\rho(E,\bk) \longrightarrow \delta(E^2-\omega_k^2)\Theta(E)
\;.\end{equation}
We then obtain for the propagator matrix
\begin{equalign}\label{fbp2}
D^{(ab)}(t,t^\prime;\bk)& = \;-\i\int\!\!dE\,\rho(E,\bk)\\
\times&  \left( (\B(n_{E\bk}))^{-1}\;
    \left(\!\!\!{\array{ll} \Theta(t-t^\prime) & \\
               &   -\Theta(t^\prime-t) \endarray}\!\!\!\right)\;
         \B(n_{E\bk})\td \,\ee{-\ie E(t-t^\prime)}\right.\\
+&
    \left.\td\B^T(n_{E\bk})\;
    \left(\!\!\!{\array{ll} \Theta(t^\prime-t) & \\
               &   -\Theta(t-t^\prime) \endarray}\!\!\!\right)\;
         (\B^T(n_{E\bk}))^{-1} \,\ee{\ie E(t-t^\prime)}\right)
\;.\end{equalign}
This is a straightforward generalization of the result
from ref. \cite{UY92c}. In the free-particle limit,
the textbook result for the finite temperature boson propagator is
recovered.

The above expression is still somewhat unsatisfactory, since the
$\B$-matrices are subject to an energy integration.
However, because of the special form \fmref{lc} we chose for the
parametrization, the integrand is {\it linear} in the
parameter $n_{E\bk}$. Thus the integration can be carried out
if one defines
\begin{equalign}\label{nbdef}
\bar{N}(t,t^\prime) & =&  \frac{1}{Z(t,t^\prime)}\,
                   \int\!\!dE\,\rho(E,\bk)\,n_{E\bk}
       \,\left(\ee{-\ie E(t-t^\prime)}+\ee{\ie E(t-t^\prime)}\right)\\
Z(t,t^\prime) & =& \int\!\!dE\,\rho(E,\bk)
       \,\left(\ee{-\ie E(t-t^\prime)}+\ee{\ie E(t-t^\prime)}\right)
\;.\end{equalign}
Some elementary matrix operations then lead to the result
for the propapagator \cite{UY92c,HU92}
\begin{equalign}\label{bpfu}
&D^{(ab)}(t,t^\prime;\bk)\;\\
=&-\i\,Z(t,t^\prime)\,(\B(\bar{N}(t,t^\prime)))^{-1}\;
    \left(\!\!\!{\array{ll} \Theta(t-t^\prime) & \\
               &   -\Theta(t^\prime-t) \endarray}\!\!\!\right)\;
         \B(\bar{N}(t,t^\prime))\td \\
&-\i\,Z^\star(t,t^\prime)\,\td\,\B^T(\bar{N}(t,t^\prime))\;
    \left(\!\!\!{\array{ll} \Theta(t^\prime-t) & \\
               &   -\Theta(t-t^\prime) \endarray}\!\!\!\right)\;
         (\B^T(\bar{N}(t,t^\prime)))^{-1}
\;.\end{equalign}
Here we have kept positive and negative energy states
separate: the inner propagator matrices are diagonal in this case.
One can, however, also
combine the two parts into one triangular
inner matrix, sandwiched between two Bogoliubov matrices:
\begin{equalign}\label{bpfu2}
&D^{(ab)}(t,t^\prime;\bk)\;\\
=&(\B(\bar{N}(t,t^\prime)))^{-1}\;\times\\
&\left(\!\!\!{\array{rr}
  -\i\Theta(t-t^\prime)\left( Z(t,t^\prime)-Z^\star(t,t^\prime)\right)
&\i\left(1+2\bar{N}(t,t^\prime)\right)Z^\star(t,t^\prime) \\
&\i\Theta(t^\prime-t)\left( Z(t,t^\prime)-Z^\star(t,t^\prime)\right)
              \endarray}\!\!\!\right)\\
&\times\;        \B(\bar{N}(t,t^\prime))\td
\;.\end{equalign}
The physical relevance of the function $\bar{N}(t,t^\prime)$
diagonalizing the propagator
becomes obvious, when we consider its equal time limit.
It approaches a constant then,
\begin{equation}\label{teq}
\lim_{t^\prime\rightarrow t}   \bar{N}(t,t^\prime)   =   N^H_k
\;\;\;\;\;\;\;
\lim_{t^\prime\rightarrow t}
        \frac{\partial}{\partial t} \bar{N}(t,t^\prime)   =   0
\;.\end{equation}
Comparision to \fmref{sdef}
gives
\begin{equation}\label{mbh}
N^H_k =  \frac{\displaystyle
                   \int\!\!dE\,\rho(E,\bk)\,n_{E\bk}}{
               \displaystyle \int\!\!dE\,\rho(E,\bk)}
=\frac{\displaystyle \lim_{t^\prime\rightarrow t}\int\!d^3\bk\,
\LGV \a_k^\dagger(t) \a_{k^\prime}(t^\prime)\RGV}{
               \displaystyle \int\!\!dE\,\rho(E,\bk)}
\;.\end{equation}
In other terms, $N^H_k$ is the time-independent equilibrium Heisenberg
density of the physical particles with momentum $\bk$.
The quantity $\bar{N}(t,t^\prime)$ therefore is the {\it observable}
fluctuating particle number of these modes.

The
separate diagonality of the inner matrices of the propagator,
i.e. the requirement of unperturbed statistical
quasi-particle propagation, is therefore equivalent to chosing
the correct physical Bogoliubov parameter.
%%%%%%%%%%%%%%%%%%%%%%%%%%%%%%%%%%%%%%%%%%%%%%%%%%%%%%%%%%%%%%%%%%%%%%%%%%%%%%%
\section{Conclusions}
The concept of quasi-particles in statistical physics is known for some
time \cite{BD71}, and it is also known that a linear
relation between the matrix elements of an interacting
propagator can be used to bring it to a triangular $2\times 2$
matrix form \cite{RS86}. We have put these two things
together and introduced statistical quasi-particles into
Thermo Field Dynamics. This leads to a
diagonal propagator for nonrelativistic models \cite{YUNA92}.

When negative energy states are taken into account, the full
propagator can also be written as diagonal matrix sandwiched
among Bogoliubov matrices, but separately so for particle and
anti-particle states \cite{HU92}.  Their combination then gives a
triangular inner propagator matrix, see eqn. \fmref{bpfu2}. We
find, that the Bogoliubov transformation necessary for this
diagonalization is given in terms of a physical parameter, the
observable fluctuating particle density $\bar{N}(t,t^\prime)$.
This diagonalization also defines stable physical modes with
fixed momentum and an average energy \cite{uh93}.

While this is clearly a conceptual advantage, the diagonalization
of the full propagator at finite temperature also has
a tremendous technical advantage over the CTP formalism.
In effect our method separates the statistical information
about the system (boundary conditions, thermal particle-hole
excitation etc.) from the purely spectral information contained in
the weight function $\rho(E,\bk)$. Let us note, that the
same separation can be achieved for non-equilibrium systems.

The application of the quasi-particle concept to a given system
not only demonstrates this technical advantage, but lends a
physical interpretation of the $2\times 2$ matrix structure of
thermal quantum theories \cite{K89,E90}. For brevity we only
state the results:  Requiring, that the triangular propagator
\fmref{bpfu} solves the diagonal components of a Schwinger-Dyson
equation gives $\rho(E,\bk)$ as function of real and
imaginary part of a retarded self energy function.

The off-diagonal component of the Schwinger-Dyson equation
contains the statistical information,
i.e. it is a consistency criterion for the function
$\bar{N}(t,t^\prime)$. For the time-independent case considered
here, we were able to derive this consistency criterion as the
condition of global thermal equilibrium. For non-equilibrium
systems, where a similar separation of statistical and
''spectral'' information can be obtained, the diagonalization condition
for the propagator is nothing but a transport
equation \cite{HU92}.
%%%%%%%%%%%%%%%%%%%%%%%%%%%%%%%%%%%%%%%%%%%%%%%%%%%%%%%%%%%%%%%%%%%%%%%%%%%%%%%


\begin{thebibliography}{99}
\bibitem{K89}{
    Proc 1st Workshop on Thermal FT and Their Applications\\
    eds. K.L.Kowalski, N.Landsman, Ch.G. van Weert,\\
    Physica {\bf 158 A} (1989) 1}
\bibitem{E90}{{\it Thermal Field Theories},\\
    Proc 2nd Workshop on Thermal FT and Their Applications\\
    eds. H.Ezawa, T.Arimitsu and Y.Hashimoto\\
    (North Holland, Amsterdam 1991)}
\bibitem{BS75}{
    H.J.Borchers, R.N.Sen,
    Commun.Math.Phys.{\bf 21} (1975) 101}
\bibitem{SKM}{
    J.Schwinger,
    J.Math.Phys. {\bf 2} (1961) 407 ;\\
    L.V.Keldysh,
    JETP {\bf 20} (1965) 1018}
\bibitem{KB62}{
    L.Kadanoff and G.Baym,\\ {\it Quantum Statistical Mechanics}\\
    (Benjamin, Reading 1962)}
\bibitem{DP91}{
   J.E.Davis and R.J.Perry, Phys.Rev.{\bf C 43} (1991) 1893}
\bibitem{AU87}{
    T.Arimitsu and H.Umezawa,
    Prog.Theor.Phys. {\bf 77} (1987) 32 and 53}
\bibitem{Ubook}{
    H.Umezawa,
    {\it Advanced Field Theory: Micro, Macro and Thermal Physics},\\
    (American Institute of Physics, in press 1992)}
\bibitem{HHW67}{
    R.Haag, N.M.Hugenholtz and M.Winnink,\\
    Commun.Math.Phys.{\bf 5} (1967) 215}
\bibitem{L88}{
    N.P.Landsman, Ann.Phys. {\bf 186 } (1988) 141}
\bibitem{HU92}{
    P.A.Henning and H.Umezawa,
    GSI-Preprint 92-61 (1992), \\ subm. Nucl.Phys. {\bf B}}
\bibitem{BD71}{
    R.Balian and C.de Dominicis, Ann.Phys. {\bf 62} (1971) 229}
\bibitem{EHUY91}{
    T.S.Evans, I.Hardman, H.Umezawa and Y.Yamanaka,\\
    J.Math.Phys {\bf 33} (1992) 370}
\bibitem{h90ber}{
    P.A.Henning, M.Graf and F. Matth\"aus,
    Physica {\bf A 182} (1992) 489}
\bibitem{UY92c}{
    H.Umezawa and Y.Yamanaka,\\
    {\it Temporal Description of Thermal Quantum Fields},\\
    University of Alberta Preprint (1992)}
\bibitem{YUNA92}{
    Y.Yamanaka, H.Umezawa, K.Nakamura and T. Arimitsu,\\
    {\it Thermo Field Dynamics in Time Representation},\\
    University of Alberta Preprint (1992)}
\bibitem{RS86}{
    J.Rammer, H.Smith,
    Rev.Mod.Phys. {\bf 58} (1986) 323}
\bibitem{uh93}{
    H.Umezawa and P.A.Henning, in preparation}
\end{thebibliography}
\end{document}